\documentclass[twocolumn,showpacs,preprintnumbers,amsmath,amssymb]{revtex4}
\usepackage{graphicx}% Include figure files
\usepackage{dcolumn}% Align table columns on decimal point
\usepackage{bm}% bold math
\begin{document} 
%\preprint{draft}
%\title{Manuscript Title:\\with Forced Linebreak}% Force line breaks with \\
%\author{Ann  Author}
%\altaffiliation[Also at] # 203-2053 West 8th Avenue, Vancouver BC %v6j 1w4, Canada
 %{Physics Department,
  %XYZ University.}
  %Lines break automatically or can be forced with \\
%\author{Second Author}%
 %\email{Second.Author@institution.edu}
%\affiliation{%
%Authors' institution and/or address\\
%This line break forced with \textbackslash\textbackslash
%}%
%\author{Charlie Author}
% \homepage{http://www.Second.institution.edu/~Charlie.Author}

%\affiliation{
%# 203-2053 West 8th Avenue, Vancouver BC v6j 1w4, Canada

%Second institution and/or address\\
%This line break forced% with \\
%}%
%\date{} It is always %\today, today,
             %  but any date may be explicitly specified
%\textsc{\today{February 12}}
%\begin{abstract}
%An article usually includes an abstract, a concise summary of the work
%covered at length in the main body of the article. It is used for
%secondary publications and for information retrieval purposes. Valid
%PACS numbers may be entered using the \verb+\pacs{#1}+ command.
%\end{abstract}
%\pacs{Valid PACS appear here}% PACS, the Physics and Astronomy
                             % Classification Scheme.
%  %\keywords{Suggested keywords}%Use showkeys class option if keyword
                              %display desired
%\maketitle
%\section{\label{sec:level1}First-level heading:\protect\\ The line
%break was forced \lowercase{via} \textbackslash\textbackslash} 
\title{Principles of equilibrium statistical mechanics revisited: The idea of vortex 
%missed vortex
%   complementary 
energy
}
%"There are none so blind as those who generalize"
\author{V. E. Shapiro}  
%\ead{vshapiro@triumf.ca}
\email{vshapiro@triumf.ca}
%\today
\affiliation{203-2053 West 8th Ave., Vancouver BC v6j 1w4, Canada}
%4004 Wesbrook Mall,
%TRIUMF, Vancouver, BC, Canada}  
\date{\today}    
\begin{abstract}
 We show that the  law of energy conservation with 
 the fact of matter stability  imply the existence 
of energy   complementary to that given by the 
function of states of interacting systems and treated, with 
the environment,  the function of states of interacting 
extended systems. 
 The complementary energy, we called it vortex, 
is  integral, not quantized, 
and causes trends  
contrary to that prescribed by  
equilibrium statistical mechanics. 
We formulate 
its principles and theorems, and
question 
traditional insights in 
thermodynamics, entropy law,
 phase transitions, persistent   
currents, Brownian motion.  
 
\end{abstract}
\maketitle  
\subsection*{Introduction}
 The physics  of phenomena is  chiefly perceived 
through the interactions  given by the energy function
in line with the
principles of holonomic mechanics and equilibrium 
thermodynamics.
This brilliantly unifying guideline created 
by Euler and Lagrange has found ways into all
pores of physics, and interpretations have  spread 
out as if 
the guideline is a genuine universal law of nature.  
 But with no outlined borders of 
validity,  the law is 
 a  default belief, a source of  circular theories  
and fallacies.

In this regard, it is worth recalling the forces 
called circulatory or vortex 
with all their cumulative impact beyond the energy 
function concept that can be huge, as known since 
the 19th century, and the  term ``dry water" coined 
by John von Neumann stuck to  viscosity-neglect  
hydrodynamic studies as inadequate, 
 see [1,2,3]. Also since the 19th century, e.g.  
[4,5], it was exposed in mechanics and 
other fields the invalidity
of the concept due to the reaction forces of ideal 
non-holonomy, performing no work on the system, as 
is the case of rigid bodies 
rolling without slipping on a  surface.
  Recall  also a general symmetry argument 
provoked by the $H$-theorem of Boltzmann and showing 
the  Loschmidt's   fundamental paradox [6] of 
 reversibility  on the way to conform 
 the real world with the 
 energy function concept.

Physics nowadays  in line with quantum mechanics 
claimed as more valid than 
classical mechanics, further spreads  the  
conviction  in the genuine energy law with no 
limits of validity, and quantum theory  has a 
commanding influence
on both fundamental and applied research. 
This common trend has various sophisticated 
possibilities to fall into the  
trap of circular theories, 
which in my judgment, 
occurs mainly due to playing with concepts of 
entropy and energy.

% How it stands, however, 
The gist of it 
%   thing 
is both energy and entropy are then
conceived as conformed to equilibrium statistical 
mechanics and thermodynamics, which 
binds them as duals of Legendre transform. 
This work aims to break the cycle and 
show 
the existence of non-conventional energy stored 
in systems in equilibrium. 
 Such energy is  not a function of 
 system states. It is related to 
 non-Hamiltonian dynamics but 
  at variance with entropy trends  and its   
 law of conservation.    
It leads to rethinking    
% change of energy paradigm of physics,     
 many standard views.  
%this with Vlasov [20]  
The work clarifies my earlier results [7,8] 
on the idea of energy  suggested first 
in connection with  the strong vortex  effect 
of high frequency fields. 
 
\subsection*{The dilemma of energy function concept
 and the issue of equal footing}
The established consistent pattern of the world 
around us is basically relaxation  to 
recurrent trends of motion. Its stability implies 
the ubiquity of  irreversible forces as the 
generalized forces whose infinitesimal work  
depends on the 
 path  of system  motion rather than just its 
 instant state.   
 
 It might seem  correct to refer the 
 irreversible forcing to the
averaging of irrelevant variables of a 
conservative many-body 
system given by a microscopic Hamiltonian and 
random initial conditions. This cue, however,     
misleads in the question of both statistical and 
dynamical 
(over fast motion) averaging  in  that 
there is no way  to come to the irreversible 
behaviors 
 from the formalism of energy functions unless 
 resorting to the inexact reasoning residing in the 
  averaging methods and 
truncations irreducible to the separation by 
canonical transformations [7,9]. 

At the same time, the perception of myriad outer 
influences, even treated as time-varying Hamiltonian 
interactions, is inevitably via smoothing which 
barely complies with the exact 
separation given by  canonical transformations, 
 hence, contributes to the irreversible  forcing 
that can greatly accumulate for long times  
along with arbitrariness in  modeling the trends. 
This is like having your cake and eating it too.
On the one hand,   the irreversible forces, unlike 
reversible, cannot be derived from a Hamiltonian or 
effective potential. On the other, insofar as  the 
true physics of phenomena is 
perceived through the interactions given by energy 
functions, 
so should be the physics of irreversible phenomena.

This dilemma 
is inherent 
to the perception in terms of energy function and 
brings in  
fundamental inexactness. 
There is no other way to account for the inexactness 
but to integrate reasoning in such terms  
with a tentative (statistical) measure of 
energy blur/relaxation 
rates. This element pertains to both 
classical and quantum mechanical descriptions. The 
uncertainty 
principle of the latter as
 related to the postulated discreteness 
of energy transfer has nothing to do with the dilemma,
and the integration in point
puts both descriptions on an equal footing.

Many phenomena in  radiation, superconductivity 
and other fields are  commonly referred to as 
indescribable classically, which might be proper 
within some specific context.   As for unconstrained 
assertions, it  contradicts the above argument
of equal footing. The same should concern 
the ideas of quantum computing claimed to be beyond 
classical physics.
Even for such phenomena as extremely deep 
cooling of matter by high frequency resonance fields, 
both ways have led to its independent prediction and 
showed    the classical way direct, free of 
 linkage to the uncertainty principle defined by 
 Planck's constant as for the cooling mechanics 
 as its limitations, see [10] and our earlier work 
 cited there. 

 Our issue here is not only 
that the physics of phenomena can be perceived 
through any self-sufficient 
construction 
and that one can't see through its wall unless 
allowing for a dual  of the 
formalism with respect to wider frameworks. It is 
also that the idea  of energy in physics incorporates 
the energy function concept as 
its part  and that it exceeds the part enormously, 
as shown below. 
Being vast it floats over the issues of equal 
footing and perception dilemma. 
Let us first outline  the existence domain 
of energy function concept. 

%Just the same the part in point is a basic element 
%for measurements. 
%Let us first proceed to its outline.   
 
\subsection*{The entrainment theorem}
Let us think of the energy function concept in terms  
of generalized thermodynamic potential 
 commonly accepted in the study of phase  transitions, 
transport through barriers and many other things. 
 The generalized potential 
of a system relaxing in steady 
conditions to a density distribution 
$\rho_{\mathrm{st}}$ connects to it by
\begin{equation}
\rho_{\mathrm{st}}(z)=Ne^{-\Phi(z)}, \quad N^{-1}=
\int  e^{-\Phi}d\Gamma
\end{equation}  %      (1)
where 
the integral is over the volume  $\Gamma$ of system 
phase space variables $z$ and the reversible motion is 
on surfaces
\begin{equation}
\Phi(z)=const.
 \end{equation}  %      (2)
The properties of the system mainly depend then on the 
local properties of the minima of $\Phi$. Also, it 
gives  insight from the observed symmetries of a 
physical system. An  analogous approach
to systems under high frequency fields is in terms of 
the picture where the hf field looks fixed or its 
effect is time-averaged. In all this, Eq.~(1) can be 
viewed as merely redefining  the 
distribution $\rho_{\mathrm{st}}$ in terms of 
function $\Phi$, whereas, taking this 
function as the energy integral of reversible motion 
provides the physical basis of the theory,
but implies rigid constraints.

Commonly, going back to  Boltzmann, Onsager, 
Graham and Haken [11], to mention a few, 
the constraints are reasoned based on microscopical 
reversibility. It corresponds to detailed balance of
transition probabilities between each pair of system 
states in equilibrium  
within the framework of autonomous Fokker-Planck 
equations treated by means of division 
of the variables and parameters into odd and even with 
respect  to time reversal, with  a  reserve on factors 
like magnetic field. In so doing the logic of time 
reversal is model-bound, ill-suits unsteady conditions 
and  the reserve rule is imposed as if universal while 
it can be easily broken, e.g., in nuclear processes 
and  where  spin-orbit interactions are a factor, 
particularly near surfaces, interfaces, dislocations. 
It makes detailed balance a non-self-maintained  
concept.

A different approach to outlining the overall domain 
of exactness in question
was suggested in [7] and will be developed here. 
Its basis  is in keeping with invariance 
under transformations of variables. On doing so 
the energy integral of reversible motion  implies the 
invariance under univalent transformations 
$z\rightarrow Z$, of Jacobian
\begin{equation}
%J=
|\det \{\partial Z_k(z,t)/\partial z_i \}|=1
\end{equation} 	%	(3)
where $i,k$ run through all components of $z$ 
and $Z$. $\Phi(z)$ (1) satisfies this condition, for
then not only  $\rho d\Gamma$ is invariant 
(being a number) 
but also $d\Gamma$. The environment as a 
fluctuation/dissipation  source for 
 the system brings in another invariance.
Connecting $\Phi$ to the system's energy 
function  implies scaling this function in 
terms of environmental-noise energy 
levels. The energy scales set this way must vary 
proportionally with the energy function in 
arbitrary moving frames $Z=Z(z,t)$ to hold $\Phi$ 
invariant.
Since the energy function changes in moving frames,
this constraint can hold only for the systems  
\emph{entrained} -- carried along on the average 
at any instant for every system's degree of 
freedom with the environment 
causing irreversible drift and diffusion.

Also account must be taken where the limit of weak 
background noise  poses as a structure peculiarity
-- transition to modeling of evolution without 
regard to  diffusion. The entrainment constraint then 
keeps  its sense as the weak irreversible-drift limit  
grasped  via the scenarios of motion along the 
isolated paths of motion  in line with 
d'Alembert-Lagrange variational principle. This  
principle  still allows 
for the ideal non-holonomic constraints that 
do not perform work on the system but reduce the 
number of its degrees of freedom, which violates  
 the desired invariance of  $\Phi(z)$. Hence, the 
invariance necessitates the domain of  entrainment  
free of that, termed ideal below.

We have discussed $\Phi(z)$ (1), but the reasoning 
holds
for any one-to-one function of  $\rho_{\mathrm{st}}$.
For the   systems describable by  
a time-dependent density distribution $\rho(z,t)$,
the adequacy of energy function formalism also
requires the entrainment ideal.
The arguments used above for the systems of steady 
$\rho_{\mathrm{st}}(z)$ become applicable there with
univalent transformations of $\rho(z,t)$ into 
$t$-independent distribution functions.  

The converse is also true:
the behaviors governed by a dressed Hamiltonian 
$\mathrm{H}(z,t)$ 
imply the entrainment ideal and the existence of
a  density distribution $\rho(z,t)$.
The  velocity function  $\dot{z}= \dot{z}(z,t)$ 
of underlying motion is then constrained by 
$\dot{z}=[z,\mathrm{H}]$ with $[,]$  a Poisson 
bracket, so the divergence $div\hspace{1pt}\dot{z}=div[z,\mathrm{H}]= 0$ 
and $div (\dot{z} f)=-[\mathrm{H},f]$ for any 
smooth $f(z,t)$.  It implies 
 \begin{equation}
\partial \rho/\partial t =
[\mathrm{H},\rho]
\end{equation}  	%	(4)
which determines $\rho(z,t)$ from a given initial 
distribution and the boundary conditions at $\vert z\vert\rightarrow\infty$  taken natural 
($\rho$ and its derivatives vanish) to preserve
the normalization and continuity, for all other 
constraints are embodied in $\mathrm{H}$. In no way 
does the solution to (4)
ceases to exist as unique, non-negative
 and not normalizable  over the phase space of $z$
where $\mathrm{H}(z,t)$ governs the behaviors.
 The entrainment ideal there takes  place since
the solution  turns into a function $\rho(\mathrm{H})$ 
in the interaction picture where $\mathrm{H}$ is 
$t$-independent. This completes the proof. 

Thus, the necessary and sufficient conditions where 
the energy function concept is duly adequate to the 
evolution described by distribution functions  come 
down to the entrainment ideal. This theorem 
lays down the overall domain of desired energy 
function adequacy. It includes  the systems  
isolated or in thermodynamic equilibrium, 
as well as entrained in steady or unsteady 
environments generally of non-uniform temperature or 
indescribable in temperature terms so long as the 
diffusion, irreversible drift and ideal nonholonomy 
can be neglected. Its criterion as an asymptotic 
limit in the parameter space of modeling is 
related to a boundary layer and 
intermittency where the limit trend can be
deprived of evidential force in the close vicinity 
of the ideal like transitions to turbulence 
for large Reynolds numbers. 

 \subsection*{
The energy measure formulation 
 }
Let us refine our concepts.
The issues of energy and energy function 
under study relate to the  systems that interact 
with the environment whose influences of short 
correlation time are accounted for 
 via the notion of entrainment introduced above.
The systems  are assumed to be describable  by a 
smooth evolution of the density distribution 
$\rho(z,t)$ of phase space  states $z$,
a set  of continuous variables $z=(x,p)$
with the generalized coordinates $x=(x_1,\ldots x_n)$ 
and conjugated moments $p=(p_1,\ldots p_n)$ of proper 
$n$ taken so in neglect of the constraints breaking 
the energy function formalism;  $z$ may include
sets of normal mode amplitudes of waves in media. The  
 smoothness  of $\rho$  will be understood to mean 
\begin{equation}
\partial \rho/\partial t=-div (v\rho)
\end{equation}  %	(5)
with $v\rho$  the  $2n$-vector flux of
phase fluid  at $z$, $t$.  Eq.~(5) 
turns into the evolution equation  of $\rho(z,t)$ 
with  $v$ treated  as a non-anticipating functional 
of  $\rho(z,t)$  that accounts for all  constraints 
on the phase flows under the boundary conditions 
taken natural for the components of $z$ set unbound. 
In neglect of nonlocal and retarded constraints, 
 $v$ is generally a $t$-dependent field
divergent in $z$. 

At that, while the $x$, $p$ of $\rho(z,t)$ of (5) is 
a set of phase space variables, the principle of 
virtual work on the system and 
the law of energy conservation, which are to be taken 
as prime as so  the material world is perceived, are 
formulated in terms of isolated paths with $x$ and 
$p$ the functions of  $t$. The notion of integration 
along the paths, hence, the work along them is 
ambiguous in conditions of diffusion; 
the mechanics pertains then solely to the forces 
of drift, which is to $v$ a vector function 
$v(z,t)$. As for a general case, we  
treat any conceivable isolated paths as 
abstraction  of the Cauchy problem of kinetics of 
$\rho$,  so the integrable correspondence  between 
the two descriptions is  to imply  the principles 
of continuity and causality. Integrable  is in the 
sense of averaging and truncations in the limit of 
short correlation time influences as defined 
in line with the basics of stochastic integral 
measure  extended in [12--14]. 

The  $n$ components
$(v_{n+1}, v_{n+2},\ldots v_{2n})$ of the 
\emph{actual} (from given initial conditions)  
phase flux at $z$, $t$ act then as
 the generalized force conjugated to 
 $x=(z_1,\ldots z_n)$, and the  scalar product
\begin{equation}
v_{n+1}\delta z_1+ v_{n+2}\delta z_2+
\ldots v_{2n}\delta z_n
\end{equation}	% (6)
represents the virtual work  on the system
irrespective of whether  this sum  is reducible 
to the variation of a scalar function or not.  
Accordingly, for the generalized coordinates taken 
 in the geometric conditions not involving time 
explicitly, the  density power on the  phase fluid 
comes down to the scalar product
\begin{equation}
(v_{n+1}v_1+\dots  v_{2n}v_n)\rho.
\end{equation}	% (7)
In particular, the energy of the  system is 
conserved as long as the integral of this density 
 over the whole phase volume remains zero,
\begin{equation}
\int(v_{n+1}v_1+\dots  v_{2n}v_n)\rho d\Gamma=0. 
\end{equation} %	(8)
 This criterion itself  bears no relation
to the entrainment ideal and shows up in both 
entrained and non-entrained systems and also as 
under steady constraints (autonomous Eq.~(5)) as 
unsteady.

Where the energy of system is conserved, there 
its energy measure exists in strict sense. So, the 
conditions where criterion (8) holds   
outlines the existence domain of the energy measure.
It includes the whole existence domain of the energy 
measure in the entrainment ideal, which is obviously
 where $v$ is a $t$-independent divergent-free 
 function of $z$, but can extend fairly far beyond 
 it -- however  far in principle  both the 
irreversible drift and  diffusion terms of $v$ permit 
and whether they are retarded and  $t$-dependent 
or   $t$-independent. 

Thereby  the  energy function
concept serves for the energy concept
 as means, instruments in  
modeling via kinetic 
equations and measurements, e.g., 
 yardstick in calorimetry, but at issue is,   
as usual, how we  interpret  the results of 
measurements and what concept is more consistent 
and wider applicable without crutches.

\subsection*{
The  canonic invariance theorem of kinetic operators 
 }
 For the evolution of $\rho$ modeled by a kinetic 
 equation 
\begin{equation}
\partial \rho/\partial t=
[\mathrm{H},\rho] + I
\end{equation}  %	(9)
where $\mathrm{H}=\mathrm{H}(z,t)$ is, unlike in 
Eq.~(4), an arbitrary smooth  function, we get from 
(5) for the term $I$
\begin{equation}
I=-div[(v-\dot{z})\rho]
\end{equation} 	%(10)
with $\dot{z}=[z,\mathrm{H}]$ the local velocity of 
Hamiltonian phase flows governed by  $\mathrm{H}$.
An important feature of presentation (9) is the 
canonical invariance of $I$  holds 
as in as off the entrainment ideal. To prove, 
note  that a canonical (univalent) transformation 
$z\rightarrow Z$ implies not only the invariance of 
$\rho$ and Poisson brackets but also the constraint 
\begin{equation}
\partial Z(z,t){/}\partial t=[Z,G]
\end{equation}	% (11) 
with $G$ a scalar function of $z,t$. Herein 
$\partial Z(z,t){/}\partial t$ is the relative 
velocity of reference frame $Z$ at $z$ and $t$, so 
the function $G(z,t)$ plays the role  of Hamiltonian 
governing this relative motion. 
%It follows 
The canonical invariance of 
$\partial \rho/\partial t-[\mathrm{H},\rho]$ in (9)
follows and, hence, of the $I$ term whatever its 
functional form may be. This formulation
generalizes our theorem IV in [7]. 
    
In the entrainment ideal, $I$ reduces to
 a  $[\mathrm{H},\rho]$-like Poisson bracket since 
the  evolution is then to be governed by a dressed 
Hamiltonian. An example is when the $I$ term is
modeled as a heat bath -- a superposition of 
Hamiltonian subsystems with randomly distributed 
initial conditions. Beyond the ideal, however, the 
entrainment theorem implies that $I$ is 
 not reducible to a $[\mathrm{H},\rho]$-like bracket, 
hence, both  canonical  invariants, 
$[\mathrm{H},\rho]$ and $I$, then cease to be 
 invariant in  the process of actual evolution  for 
 any choice of $\mathrm{H}(z,t)$. 
 
Abstracting of  the evolution,  the state of $\rho$ 
at any given instant $t=t_i$ can be taken 
for ideally entrained by fitting. Due to this and since 
$\rho$ is assumed smooth in $t$, the effect of the 
irreducibility of $I$ is weak for  $t$'s  close to 
 $t_i$, so it might seem reasonable to judge about 
its figure of merit by  popular perturbation methods 
of celestial mechanics, e.g. [15].  But such insight 
is insufficient. It fails in the long run  beyond 
the ideal entrainment to match the future with 
the past and so conforms  to the trends of $\rho$  
in line with a dressed Hamiltonian, which
conduces to the belief in this theory  beyond its 
above-established rigid 
constraints.  

Physically, as the canonical transform is equivalent 
to the imposition of fields given by Hamiltonian 
$G(z,t)$, the fields superimposed on the system 
affect directly the conditions of its entrainment, 
and so the reversibility of the overall evolution of 
$\rho$ is affected in response. 
This generally translates into a vortex (in spatial 
subspace) impact exerted on the system  in the 
picture at ``rest", where the field $G$ looks frozen. 
Though small at $t\rightarrow t_i$, it tends to 
accumulate exponentially and is not a nuisance. 
In particular, this shows up vigorously  
for the systems in high frequency resonance fields, 
especially at parametric and combination resonances, 
as elucidated in [7] and our earlier work cited 
there. 
The arising steady states of $\rho$ and behavior 
near them in the picture at rest were shown 
 to differ radically from that given by
the theory of generalized thermodynamic potential.
 
\subsection*
{The energy duality 
} 
 Let us focus on  the systems relaxing to  stable distributions of their  states in stationary conditions.    
%in stationary conditions. 
The energy-measure criterion (8) includes then 
the whole  area of reversible-motion criterion (2) but 
is not confined to it at all, which is indicative of the fact  
that the conditions of $\rho_\mathrm{st}$ where the 
conserved energy is indescribable 
via a  generalized thermodynamic potential are common 
and may range far. In terms of Eq.~(9) we get
\begin{equation} 
[\mathrm{H},\rho]+I=0
\end{equation} 	%	(12) 
where the branch $I$  acts on a par with  
$[\mathrm{H},\rho]$ in jointly keeping the 
circulation and  transformations of conserved energy 
both within and beyond the entrainment ideal. 
As for beyond, it implies  the conserved energy 
irreducible to a function of system states. 
Accordingly, whereas  the conserved 
energy of motion (chaotic motion including) can be 
conceived within the ideal as the circulation of the 
 kinetic and potential energies  within the 
framework of  dressed Hamiltonian, 
 the conserved energy circulating in the 
systems beyond the ideal  includes or constitutes 
entirely the energy form complementary 
to the forms describable by a  
Hamiltonian, hence, quantizable. 

Such energy, which we
called  integral or vortex energy, is 
also under no bound to the principles of  detailed 
balance, energy transfer directionality, stability and  
 preference of phases   
 -- all that given by the 
conventional theory of phase transitions, transport 
through barriers and other phenomena    
based on the generalized thermodynamic potential.  
The  dualism associated with the
 complementary energy in point also
has nothing to do with 
the particle-wave dualism in quantum mechanics  
 and the concept of energy transfer by energy 
quanta. It questions the all-physics 
adequacy of quantum approach.
The quantum approach, just as the classical one, 
to be adequate would imply incorporating the scope  
of energy - energy function duality.  

 \subsection*
{The vortex energy and the directional Brownian 
motion   }
 Look first at a particle hopping upon a horizontal 
reflecting plate.  Gravity tends to bring it  into 
contact with the plate and the ambient noise  keeps 
it hopping in stationary conditions. Now let the 
particle be charged and the field of permanent 
magnet be applied horizontally.
This causes the hoping particle to drift 
in the direction across the field. The net drift 
is modified but does not vanish when the 
reflecting surface is uneven or rolled or forms a  
box, and it persists as the particle motion  
state relaxes  to a stationary $\rho_\mathrm{st}$. 
The same trend is for 
a number  of interacting  charged particles 
in the presence of reflecting walls. 
 The energy of steady macromotion   is then 
conserved,  but it is not describable by an 
energy function of macromotion states and   
holds vortex energy.  
The general theorem shown below makes it evident. 
 
The existence of distribution $\rho_\mathrm{st}(z)$ 
in a stable entrainment ideal in stationary 
conditions implies, along with relaxation to the 
ideal, the system's dressed Hamiltonian 
$\mathrm{H}(z)$ to exist, be bound below,  
commutate with $\Phi(z)$ and be a monotonic 
function  of $\Phi$. Thereat, the vanishing 
irreversible forcing on the average for every 
component $i$ of system variables $z$ implies 
according to (9) and (12) the constraints 
\begin{equation}
\left(f_i -d_{ik}\frac{\partial}{\partial z_k}+ 
\ldots \right)\rho_\mathrm{st}(z)=0
\end{equation}	%	(13)
where $f=\{f_i(z)\}$ is the irreversible drift forces, 
 $d=\{ d_{ik}(z)\}$ is
a symmetric non-negative definite matrix of diffusion 
and ellipsis  stands for the higher order diffusion 
terms of expansion  of $I$ into a series in 
$\partial/\partial z$. As $I$ is generally an 
integrodifferential form in $z$, so is 
the operator bracket  of (13). 
The constraints of (13)  generalize 
the conditions of detailed balance. 

Neglecting the higher order  terms in the bracket  
reduces  Eq.~(13)  to the  algebraic 
fluctuation-dissipation relations  
\begin{equation}
f_i=-(d\Phi /d \mathrm {H})d_{ik}\dot{z}_k 
\end{equation} 	%	(14)
with $\dot{z}=[z,\mathrm{H}]$.
For the distribution $\rho_\mathrm{st}$ of 
Maxwell-Boltzmann form  and  general  Gibbs form, 
$d\Phi /d \mathrm{H}=\beta $ is independent of 
$\mathrm{H}$, which reduces (14) to 
\begin{equation}
f=
-\beta d \dot{z}=-\beta d [z,\mathrm{H}].
\end{equation} 		%	(15)
$\beta^{-1}=\Theta$ is the energy scale  
of absolute temperature whose meaning  
expounds the known 
equipartition theorem: 
for every component of $z$ 
(coordinate or momentum) whose contribution to 
$\mathrm{H}$ 
reduces to a square term, say, $k_1(z_j-k_2)^2$ 
with  $k_1>0$ and $k_{1,2}$ independent of $z_j$ but 
may  depend on other components of $z$ and $t$,   
its mean over the Gibbs statistics comes to 
$\langle k_1(z_j-k_2)^2\rangle=\Theta$.

It is easily seen that the $\rho_\mathrm{st}$ taken 
a Gibbs rules out persistent currents since for any 
$\langle\dot{z_i}\rangle $, a function of $z_i$ 
averaged over the phase subspace  off $z_i$, one  
gets on integrating by parts 
\begin{equation}
\langle\dot{z_i}\rangle=N\int [z_i,\mathrm{H}]e^{-
\beta\mathrm{H}}(d\Gamma/d z_i) =0 
\end{equation}	%	(16)
by virtue of natural boundary conditions for $\Phi(z)$. 
The theorem  $\langle\dot{z_i}\rangle=0$ holds 
 not only for Gibbs but for any arbitrary statistics 
 of $\rho_\mathrm{st}$,  a function of $z$ via 
$\mathrm{H}(z)$.  The proof ensues
  from  $[z,\mathrm{H}]=[z,\Phi]\mathrm{H}'$ with 
$\mathrm{H}'=d\mathrm{H}/d\Phi>0$, 
for the  sign of every $[z_i,\Phi]$ is implied so 
for stability.  

These results show  no place for a stable macromotion 
state in stationary conditions within the framework 
of generalized thermodynamic potential.   
Such states  are thus a Litmus test of conserved 
vortex energy. A distinctive feature of the phenomenon 
is robustness as the stability of macromotion state 
is to be asymptotic, with relaxation a factor 
and with reversion in response to weak perturbations. 
It extends the paradigm of Brownian motion caused by 
eternal chaos as non-directional to that of 
directional motions caused by eternal chaos. 
 
While any system at a certain standing can be taken
via fitting as ideally entrained, 
governed by an energy function of its states, 
the theories of transition from there  under 
a shift of parameters to a stable macromotion 
beg a question whenever the emerging macromotion 
state is again treated as a state given by an energy 
function. The macromotion  is then attributed to 
spontaneous symmetry breaking, topological defects 
and what-not, which is problematic as it implies    
the conditions (13) to be somehow miraculously 
restored. Anyhow, in the end  
one faces  
the above theorem banning 
a stable 
macromotion within  this beaten path down-the-line. 
To claim the phenomenon as just quantum  is not 
sufficient, for  as in classics this needs 
consistently applied principles to  account  for  
the transition to a stable stored  energy of vortex 
form.  

In contrast to the essence of pattern formation as    
a process  that makes the  Cauchy problem of kinetic 
equation (9) and its quasi-static 
($\partial/\partial t \rightarrow 0$, not just 
$\partial/\partial t = 0$) limit  (12) the corner 
stone of the theory of energy, as we do, the theory 
of phase transitions in question makes, in fact,    
the boundary value problem of Liouville type   
kinetic equations  the corner stone.
This results in the geometrization beauty
of kinetics but rules out the formation intrinsic 
to a stable non-entrained state, hence,
the macromotion and vortex energy. 

\subsection*{Thermodynamic laws in the light 
%from the angle 
of vortex energy
%Comment on thermodynamic laws
}
Let us look  into equilibrium thermodynamics. 
It proceeds  from the existence  of internal energy 
$E$ of  system as a function of  external parameters 
$a=\{a_k\}$ and temperature $\Theta$ so that 
 the   differential $dE$ in space $(a,\Theta)$  
\begin{equation}
dE = 
\frac{\partial E}{\partial \Theta} d\Theta 
+\frac{\partial E}{\partial a_k} d a_k 
= \delta Q+\delta W
\end{equation}
expresses the first law by introducing the   heat 
transfer $Q$  as the difference between the internal 
energy and the 
work on the system  $W$ defined for any processes as     
purely mechanical, for $\Theta$ fixed. 
For the processes to proceed the parameters  
are assumed to  vary in time, but slowly - in the 
quasi-static limit  $d(a,\Theta)/dt\rightarrow 0$.
Whereas $Q$ and  $W$    may freely depend on the 
path chosen in $(a,\Theta)$ with  $\delta Q$ and 
$\delta W$ not bound to be exact differentials,   
Eq.~(17) implies for any cyclic process 
\begin{equation}
\oint \delta Q=-\oint\delta W . 
\end{equation}	% (18)  
Therein lays the principle of equivalence between the 
work and  heat. Being for any path in ($a,\Theta$), 
it means two separate relations of detailed  energy 
balance, Eq.~(18) for the work of irreversible forcing 
and Eq.~(17) with $\delta Q+\delta W$ replaced by  
their reversible part for the reversible forcing. 
Treating both as a projection of the separating 
principle  between the balances of reversible and 
irreversible forcing we formulated in the paragraph 
with Eq.~(13)  shows the first law as the  law of 
energy conservation bound to energy function concept  
for the case and, since  equilibrium is treated as 
a stable state, implies relaxation towards the 
minimum of energy function of system states in 
terms of $a,\Theta$ without introducing any entropy 
function.

The second law of thermodynamics in this regard
specifies  the  equation of system state, its 
caloric and thermal relations -- by  assuming the 
energy function is additive with respect  to the 
partition of system volume,  a one-dimension 
external parameter, in  independent small parts. It 
best fits the ideal gas confined by rigid walls, 
 is  in line with Gibbs statistics of 
$\rho_\mathrm{st}$ and poses  the energy $E$  and 
forces $A_k=-\partial E/\partial a_k$ as the 
averages 
\begin{equation} 
E=\int\mathrm{H}e^{(\psi-\mathrm{H})/\Theta} 
d\Gamma, 
\end{equation} 		%	(19)
\begin{equation}
A_k= \int (-\partial\mathrm{H}/\partial a_k)
e^{(\psi-\mathrm{H})/\Theta}d\Gamma
\end{equation} 		%	(20)
with
\begin{equation} 
\psi=-\Theta \ln N, \quad
N= 
\int e^{-H/\Theta}d\Gamma
\end{equation}  		%	(21) 
and the Hamiltonian $\mathrm{H}$ assumed a 
function of $z$ and slowly varying parameters $a$
but not $\Theta$. It follows 
\begin{equation}
A_k=-\frac{\partial \psi}{\partial a_k}, \quad
E=\psi-\Theta\frac{\partial \psi}
{\partial \Theta}, 
\quad S=-\frac{\partial \psi}{\partial \Theta}. 
\end{equation} 		%	(22)
The first relation shows  $\psi(a,\Theta)$ as the 
Helmholtz free energy, a  function comprising the 
work of  forces $A=\{A_k\}$, so the second shows
 $\Theta\partial \psi/\partial \Theta$ as the  
 binding energy function; and the problem of energy 
and forces at equilibrium  is  determined by a 
single function $\psi$ of system states.      $S$ 
represents  the entropy function 
$\int (\delta Q/\Theta)$ introduced in pure 
thermodynamics by postulating the existence of the 
integrating multiplier of $\delta Q$ with $1/\Theta$, 
so the constraint on function $S(a,\Theta)$ 
to be maximal at thermodynamic equilibrium implies 
the direction of relaxation only to such ideal.  
Gibbsian concept makes more sense in physics and 
shows  entropy in (22) as not a self-sustained 
notion for that matter and that  the first and 
second laws do not extend to the vortex energy and 
its trends. 

The latter assertion is to be common  to any 
extensions of entropy function 
as within the  first law (17) as for
more general entrainment ideal conditions. 
Indeed, the entropy function and   the 
generalized potential must  always commute  
since the ideal entrainment holds where this 
potential for the system is its energy integral. 
The  violation of entropy conservation law would 
mean that the entropy is not a function of 
parameters entering in the potential for the case. 
This is also so in stationary  conditions where the 
law of energy conservation holds beyond the 
entrainment ideal,  
for the opposite would then mean the existence of 
the energy integral of the system. As to the  
conservation law of entropy in conditions where the 
energy of system is not conserved,   the entropy 
cannot be related to the  system energy, for such 
notion does not exist then, which means the entropy 
conservation is out of physical perception.

Of various entropy functions linked to the 
conservation law 
$\oint (\delta Q/\Theta)=\oint \delta S=0$ for slow 
cyclic processes, only Gibbs statistics  assigns to 
the pure thermodynamics the meaning given by the 
equipartition theorem. But at that,  
only a small area of Gibbs statistics domain fits 
the thermodynamics, as particularly evident from the 
paragraph with Eqs.~(13)--(16). It implies
$\mathrm{H}$  to  be bound from below and 
 the additivity postulate to limit its long-ranged 
interactions, and the  interactions and parameters  
entering into $\mathrm{H}$  should not depend on 
$\Theta$ and  statistical factors  -- to preserve 
the very separating principle between the balances 
of reversible  and irreversible forcing and avoid 
ambiguity in its definitions. 

In this light, the known  Landau theorem [16], 
 that a closed system of interacting parts in thermal 
equilibrium admits only uniform translation and 
rotation as a whole,  referred to as the outright ban 
on classical routes to persistent currents, should 
not be treated so.  The proof [16]  proceeds from
 the system's entropy $S$ taken in the form of a sum 
$\sum S_i$ where each summand $S_i$ is a function 
of the difference $E_i-P_i^2/2m_i$ between the total  
and  kinetic energy only of part $i$. The statistics 
of $\rho_\mathrm{st}$ is not specified, 
but the additivity assumption is very restrictive. 
%Also, 
Also, once the entropy function is taken even in the 
moments $P_i$'s of system parts, so the distribution 
of $\rho_\mathrm{st}$ is,  
which automatically rules out persistent currents.  
But the general conditions of outright ban do not rely on 
the parity in point, as seen from Eq.~(16) and the 
theorem  below it. 

We now make a 
 comment on the  theory of matter stability, its 
element based 
on Gibbsian thermodynamics for Coulomb systems.  
By the rigorous  theory, see [17,18], and the  
mean-field theories going back to Debye the screening 
of long-range Coulomb potential $1/r$ between moving 
charges by the  charges of opposite sign in matter  
makes the potential short-ranged, 
so the free energy per unit volume is bound below and 
tends to a finite  limit as the system volume 
increases. But all that presumes Gibbsian 
thermodynamics. The  sufficient conditions would 
include 
the stability with respect to wider possibilities of 
energy conservation, for the screening arises due to 
the diffusion and relaxation of gradient of 
charge-particle density under field perturbations.   
Within the domain of generalized thermodynamic 
potential the sufficiency reduces to criterion (13), 
whereas  beyond, the vortex energy emerges,  the  
energy function concept loses force, so the energy 
integral transits into the energy functional (8)   
and the stability criterion (13) 
into that where $f$ comprises all drift forces, 
which is accessible for measurements.

Just as important  are   the constraints imposed on 
particle systems due to enclosure  needed for their 
confinement  as it may not comply  without  vortex 
energy.  This is so for our example of particle 
hopping in a  box and 
plausible in phenomena where surfaces,  interfaces, 
dislocations are a factor. Besides nonholonomy the 
nonconcavity of conservative field Hamiltonian may emerge.
It might concern, e.g., superconducting topological 
insulators commonly treated in quantum terms. Recall 
also the electron fluid instability 
suggested by Vlasov [19] by analogy with 
the  physics of capillary waves  going 
back to Stokes and Rayleigh [20] 
-- the attraction of surface particles to the 
bulk of fluid contributes to the 
negative potential energy of ripple wave motion, so 
such states  can evolve into a steady ripple that 
transports mass and charges. Obviously for the phenomenon 
to exist as  robust, held long compared with 
relaxation time in conditions of vanishing work on the 
system and scattering, it implies stored vortex energy 
for stabilization.

\section*{%Summary
Concluding remarks
}
The presented idea of energy as  a collective concept 
of interacting systems departs from the traditional 
insight. 

The central element of departure is  the law of 
conservation of energy, where the energy we 
proceed from 
characterizes  the ability to produce work  
defined by  d'Alembert principle, not merely its 
surogat given by Hamiltonian of systems. 
Also, while the stored energy is  
a measure 
to be given  through the evolution of 
distribution function of system states, 
taking the 
energy function concept for granted implies 
substituting the Cauchy problem of equations 
governing the evolution by a boundary value problem.
The departure is thus from the physics of  basically 
predetermined world to that of real, diverse  world
where nothing happens by itself but depends on 
circumstances. 

A  stumbling block on the path to this diversity is 
that conserved energy  being a measure born in 
mechanics, not kinetics, is tied up to the  notion of 
characteristics which is applicable only to a very  
particular type of kinetic equations.  The   
integrable correspondence between the two descriptions 
of evolution we came to based on the principles of 
continuity and causality  gets over that. 
The correspondence 
follows the line of how the measurements 
of kinetics are perceived  and appears completely 
consistent with the  canonical invariance feature 
we have formulated here  for  kinetics itself. 

These principles, together with the fact of existence 
of stable matter in stationary conditions, imply the 
existence and ubiquity of stored vortex energy 
as complementary to that prescribed by 
the energy function concept. 
The presented extension 
to this concept of common use for interpretations and 
predictions is like extension from integer numbers to 
all reals but deeper since it is on functional level. 
It implies 
the stable  self-sustained  motion states in 
equilibrium  out of generalized potential concept.  
%The two detailed balances merge there into one inseparable. 
% with both drift and diffusion.  Being cumulative, the 
% so the 
%extended energy notion incorporates both drift and diffusion. 
Essentially, the extended notion of stored energy 
is intrinsic  of non-vanishing  drift and diffusion.  
The cumulative 
effect on equilibrium state, its stability and 
fluctuations can be huge.  
 
The stored vortex energy being not a function of 
system states is integral, not quantized, and 
 appears to be under no bound to the trends of 
conventional 
 equilibrium statistical mechanics and first and 
 second laws of thermodynamics  as the steady 
 states are determined by non-selfadjoint 
 operators characteristic of indecomposability. 
The law of conservation of entropy is then at 
variance with the law of conservation of energy, 
and taking the energy as prime makes the entropy 
concept unacceptable.  So the entropy argument is 
not suited for the trends of vortex energy and its 
existence domain; the stability criterion (13) 
with $f$ comprising all drift forces is then of 
importance as accessible for measurements. 

The existence domain of stable matter  may  
extend or shrink not bound to changes of energy 
function  at all  -- the notion of energy function 
of system states and their dressed option 
loses sense  as the vortex energy emerges.  
We meet with roughly the same vortex 
energy circulation in equilibrium in the 
systems under high frequency 
fields in the picture where the hf field is frozen, 
see [7]. So a vast additional range of objects 
and phenomena has a bearing on the matter. 
All that questions the all-physics 
adequacy of pure quantum approach  as it 
has for the  classical energy function concept   
and its formulation in  relativity physics. 
In particular, it   
stands to reason that the vortex energy has a bearing on 
 black holes, dark energy and dark matter. 

An important result to note is our theorem that rules 
out any stable macromotion in stationary conditions 
as soon as the  distribution function   
$\rho_\mathrm {st}(z)$ of system states, of any 
statistics,  is treated within the framework 
of generalized thermodynamic potential.
It makes persistent currents  a Litmus test of 
vortex energy, imposes constraints on the 
traditional theory of phase transitions, 
extends the paradigm of Brownian motion caused by 
eternal chaos as non-directional to that of 
directional, and gives a natural  solution to the 
fundamental Loschmidt's paradox. 

%These common misconceptions opened up 
%The scope of 
%revealed vortex energy and the 
%common misconceptions opened with the revealed 
%vortex energy 
%lead to a new perspective of physics    

%The revealed common misconceptions associated with 
%revealed vortex energy  are reminiscent of flat Earth myths
%and call into changing the whole paradigm of physics 

The revealed vortex form of  energy 
and  common 
misconceptions of conventional  energy concept  
bring forth  the necessity to change 
the whole  paradigm of physics based 
 on energy and entropy conservation laws that is   
reminiscent of flat Earth myths. 
The change extends the horizons of search 
for new forms of energy,  matter and  
macromotion and is important for applied research. 
 
\acknowledgements
The author would like to thank Prof. Ian Affleck 
and  student Zach Sagorin of UBC  for the interest 
and  assistance in proof-reading the English of 
the work.

\end{document}